\documentclass{mn2e}
\usepackage{graphicx}
\def\go{
\mathrel{\raise.3ex\hbox{$>$}\mkern-14mu\lower0.6ex\hbox{$\sim$}}
}
\def\lo{
\mathrel{\raise.3ex\hbox{$<$}\mkern-14mu\lower0.6ex\hbox{$\sim$}}
}
\def\simeq{
\mathrel{\raise.3ex\hbox{$\sim$}\mkern-14mu\lower0.4ex\hbox{$-$}}
}

\def\etal{{\it et al.\ }}

\def\etal{{\it et al.\ }}

\def\be{\begin{equation}}
\def\ee{\end{equation}}
\def\bea{\begin{eqnarray}}
\def\eea{\end{eqnarray}}

\def\etal{{\sl et al.\ }}

\def\hw2{{\hat W}^2}
\def\go{\mathrel{\raise.3ex\hbox{$>$}\mkern-14mu
             \lower0.6ex\hbox{$\sim$}}}
\def\lo{\mathrel{\raise.3ex\hbox{$<$}\mkern-14mu
             \lower0.6ex\hbox{$\sim$}}}
\def\ltorder{\mathrel{\raise.3ex\hbox{$<$}\mkern-14mu
             \lower0.6ex\hbox{$\sim$}}}
\def\gtorder{\mathrel{\raise.3ex\hbox{$>$}\mkern-14mu
             \lower0.6ex\hbox{$\sim$}}}

\def\eps2{{\epsilon^2}}

\def\msun{{\rm M_{\odot}}}

\begin{document}

\title[An {\it XMM-Newton} observation of the Narrow Line Seyfert 1
Galaxy, Markarian 896]
{An {\it XMM-Newton} observation of the Narrow Line Seyfert 1
Galaxy, Markarian 896}
\author[K.L. Page \etal]{K.L. Page$^{1}$, P.T. O'Brien$^{1}$,
J.N. Reeves$^{1}$, A.A. Breeveld$^{2}$\\
$^{1}$ X-Ray and Observational Astronomy Group, Department of Physics \& Astronomy,  
University of Leicester, LE1 7RH, UK\\
$^{2}$ MSSL, University College London, Holmbury St. Mary, Dorking, Surrey, RH5 6NT, UK}

\date{Received ** *** 2002 / Accepted ** *** 200*}

\label{firstpage}

\maketitle

\begin{abstract}
{\it XMM-Newton} observations of the NLS1 Markarian~896 are presented.
Over the 2--10~keV band, an iron emission line, close to 6.4~keV, is seen. The line is just resolved and has an equivalent width of $\sim$~170~eV.
The broad-band spectrum is well modelled by a
power law slope of $\Gamma$~$\sim$~2.03, together with two blackbody
components to fit the soft X-ray excess. Using a more
physical two-temperature Comptonisation model, a good
fit is obtained for an input photon distribution of kT~$\sim$~60~eV and
Comptonising electron temperatures of $\sim$~0.3 and 200~keV. The soft excess
cannot be explained purely through the reprocessing of a hard X-ray
continuum by an ionised disc reflector.

\end{abstract}


\section{Introduction}
\label{sec:intro}

Narrow Line Seyfert 1 galaxies (NLS1s) are a subset of AGN with
particularly narrow Balmer lines, these being only slightly broader (H$\beta$~$\leq$~2000~km~s$^{-1}$;
Osterbrock \& Pogge 1985)
than the forbidden lines; in this respect they are similar to Seyfert
2s. However, NLS1s also show an [O{\sc iii}] to H$\beta$
ratio of $>$3, which is one of the distinguishing features of
Seyfert 1, rather than Seyfert 2, type galaxies (Shuder \& Osterbrock 1981). The
third characteristic is the frequent presence of strong Fe~{\sc ii} (or
higher ionisation iron) emission lines; these, again, are more usually
seen in Seyfert 1s than Seyfert 2s.

Observations with {\it ROSAT} found that
NLS1s tend to have steep 0.1--2.4~keV X-ray slopes, leading to
soft luminosities which are higher than can be
explained by the reprocessing of the hard X-ray continuum. They also
have  increased
variability in this region, with doubling timescales as short as a few
minutes (e.g., Boller,
Brandt \& Fink 1996). A possible explanation for the extreme
properties of these AGN is that they contain relatively low mass black
holes, which are accreting at a high (near-Eddington) rate (Pounds,
Done \& Osborne 1995). 

In this paper, we present {\it XMM-Newton} observations of
Mrk~896 (also known as MCG$-$01$-$53$-$008 and
IRAS~F20437$-$0259). With an H$\beta$ FWHM of 1135~km~s$^{-1}$
(V$\acute{e}$ron-Cetty, V$\acute{e}$ron \& Gon\c{c}alves 2001), Mrk~896 lies
towards the centre of the range of NLS1s line-widths
($\sim$~500--2000~km~s$^{-1}$) and is a low redshift
(z~=~0.0264) object, with a 5GHz flux density of 38~mJy (Bicay \etal 1995). 
Boller \etal (1996) measure $\Gamma$ to be 2.6~$\pm$~0.1
over the {\it ROSAT} band of 0.1--2.4~keV, while it was found to be
2.82$^{+0.13}_{-0.14}$ during an earlier observation
(Boller \etal 1992).

\section{XMM-Newton Observations}
\label{sec:xmmobs}

Mrk~896 was observed twice in revolution 355 (2001 November 15), as part of
a sample of NLS1s. All instruments (EPIC -- European Photon Imaging
Camera --
Str\"{u}der \etal 2001, Turner \etal 2001; RGS -- Reflection Grating
Spectrometer --
den Herder \etal 2001;
OM -- Optical Monitor -- Mason \etal 2001) were used over the second
observation. During the first observation, the EPIC
instruments were closed; data were obtained for the RGS and OM, however. The total exposure times for the various instruments were as follows: MOS -- $\sim$~10~ks; PN -- $\sim$~8~ks; RGS -- $\sim$~20~ks; OM -- $\sim$~16~kS

The pipeline-produced EPIC event-lists were filtered using {\sc xmmselect}
within the {\sc sas} (Science Analysis Software); single- and
double-pixel events (patterns 0--4) were chosen for the PN data, while
the range 0--12 was used for the MOS cameras. Source spectra were
extracted from the images using a  circular region of 45 arcsec
radius. This same region was then moved to an adjacent area of `blank sky', to
obtain a background spectrum.

The RGS~1 spectra for the two observations were co-added, as were the
RGS~2 datasets.
The spectra were binned, using the {\sc ftool}
command {\sc grppha}, to provide a minimum of 20 counts per bin for
the EPIC spectra. 
The {\sc Xspec} v11.0.1 software package was then used to analyse the
background-subtracted EPIC and RGS spectra, using the most recent response
matrices. ({\sc rgsrmfgen} was run within the {\sc sas} to obtain
the relevant RGS responses.)

Throughout this letter, H$_{0}$~=~50~km~s$^{-1}$~Mpc$^{-1}$ and
q$_{0}$~=~0.5 are assumed; unless stated otherwise, errors are given
at the 1~$\sigma$ level.

\section{Spectral Analysis}
\label{sec:specanal}

\subsection{EPIC data}
\label{sec:specanal_epic}

As is conventional, the first model fitted to the joint MOS+PN data was
a single absorbed power law (N$_{H}$ fixed to the Galactic value of
3.89~$\times$~10$^{20}$~cm$^{-2}$; Murphy \etal 1996), over the entire 0.2--10~keV
bandpass. This provides a poor fit, with
$\chi^{2}_{\nu}$~=~1168/720, mainly due to a strong upward curvature
below $\sim$~2~keV. Constraining the model solely to the 2--10~keV band,
however, produces an excellent fit for a photon index $\Gamma$~$\sim$~1.96
($\chi^{2}_{\nu}$~=~261/266; fit 1 in Table~\ref{fits}). 

\begin{table*}
\begin{center}
\caption{} 
\label{fits}
Fits to the {\it XMM-Newton} data.\\
$^{a}$ Rest-frame energy of the emission line.
$^{b}$ Intrinsic line-width.
$^{c}$ Blackbody temperature.
$^{d}$ Break Energy of 1.08~$\pm$~0.04~keV.
$^{f}$ frozen.
\begin{tabular}{p{0.8truecm}p{1.0truecm}p{2.0truecm}p{1.8truecm}p{1.8truecm}p{2.0truecm}p{1.2truecm}p{2.0truecm}p{2.0truecm}}
\hline
Fit & Range & Model & $\Gamma$ & E$^{a}$     & $\sigma$$^{b}$ & EW   & kT$^{c}$ & $\chi^{2}$/dof\\
    & (keV) &       &          & (keV) & (keV)    & (eV) & (keV)\\
\hline
1 & 2--10 & PL & 1.96~$\pm$~0.06 & & & & & 261/266\\
&\\
2 & 2--10 & PL+GA & 2.04~$\pm$~0.04 & 6.36~$\pm$~0.06 &
    0.124~$\pm$~0.075 & 182~$\pm$~65 & & 246/263\\
&\\
3 & 2--10 & PL+DISKLINE & 2.06~$\pm$~0.04 & 6.42~$\pm$~0.12 & &
244~$^{+40}_{-115}$ & &245/263 \\
&\\
4 & 0.2--10 & PL+GA+BB & 2.16~$\pm$~0.02 & 6.36$^{f}$ & 0.124$^{f}$&
    218~$\pm$~50 & 0.088~$\pm$~0.002 & 713/717\\
&\\
5 & 0.2--10 & PL+GA+2BB & 2.03~$\pm$~0.04 & 6.36$^{f}$ & 0.124$^{f}$&
    165~$\pm$~48 & 0.084~$\pm$~0.003 & 698/715\\
 & & & & & & & 0.224~$\pm$~0.022\\
&\\
6 & 0.2--10 & BKNPL+GA & 2.65~$\pm$~0.02$^{d}$ & 6.36$^{f}$ &
0.124$^{f}$& 222~$\pm$~54 & &740/717\\
 & & & 2.15~$\pm$~0.02\\
&\\
7 & 0.2--10 & 2PL+GA & 2.94~$\pm$~0.09 & 6.36$^{f}$ &
0.124$^{f}$& 152~$\pm$~48 & &741/717\\
 & & & 1.71~$\pm$~0.10\\ 
\hline
\end{tabular}
\end{center}
\end{table*}

There were seen to be positive residuals
to the fit around 6.4~keV, in both the MOS and PN data, so a Gaussian
emission component was added. Allowing both the energy and width of
the line to go free, the best fit was obtained for a rest energy of
(6.36~$\pm$~0.06)~keV, with $\sigma$~=~(0.124~$\pm$~0.075)~keV. The
equivalent width was measured to be 182~eV (fit 2 in Table~\ref{fits}).
This improved the fit by $\Delta\chi^{2}$ of
13 for 3 degrees of freedom, corresponding to a probability of
$>$99~per~cent.  A second
Gaussian did not improve the fit significantly. It should be noted that freezing the energy to 6.4~keV, for
neutral iron emission, gave an equally acceptable fit. Also, if the line is
fixed to be unresolved ($\sigma$~=~0.01~keV), the fit is only worse by
$\Delta\chi^{2}$ of one, for one degree of freedom; this line has an
equivalent width of $\sim$~140~eV. This is consistent with lines seen
in other objects (O'Brien \etal 2001; Reeves \etal 2001; Pounds \etal
2001; Kaspi \etal 2001; Lubinski \& Zdziarski 2001). However, if we assume there is no intrinsic
narrow line, a diskline model (Fabian \etal 1989) can be tried
(fit 3); this was found to be a good fit, with an inclination angle of
24~$\pm$~15$^{o}$ for the disc (inner radius fixed at a typical value of three
Schwarzschild radii; emissivity index~=~$-$2). In this case, the narrow component corresponds to the blue `horn'
of the profile. It is, unfortunately, not feasible to decide with statistical significance between between the Gaussian and diskline models.

\begin{figure}
\begin{center}
\includegraphics[clip,width=0.7\linewidth,angle=-90]{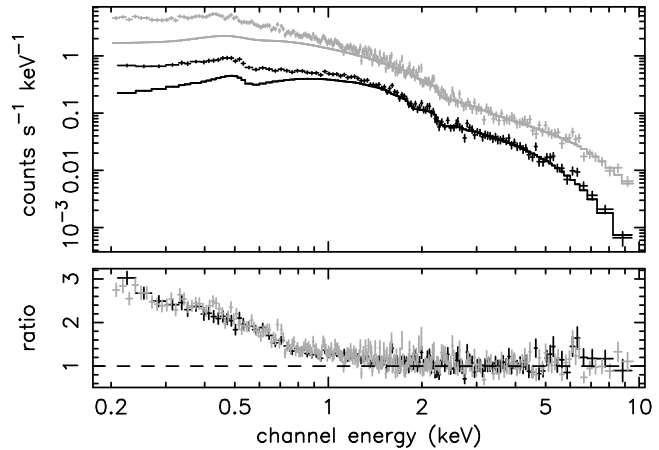}
\caption{The 0.2--10~keV MOS (black) and PN (grey) spectra of Mrk
896. The 2--10~keV power law has been extrapolated
down to lower energies, showing the presence of a strong soft excess. Positive
residuals can be seen just above 6~keV, indicating an emission line.}
\label{softexcess}
\end{center}
\end{figure}

Extrapolating the 2--10~keV power law down to 0.2~keV revealed an
obvious broad soft excess (Fig.~\ref{softexcess}). To model the
broad--band spectrum, the iron line width and energy were fixed to the
2--10~keV values and blackbody (BB) components used to parametrize the
soft excess. It was found that the breadth of the excess required two
BBs to fit the observed spectrum (fit 5 in Table~\ref{fits}).

Since the previous {\it ROSAT} papers had modelled the soft excess
region using a power law, it was decided to try fitting the {\it XMM}
soft X-ray spectrum with a simple power law. Over the 0.1--2.4~keV
{\it ROSAT} band-pass we get $\Gamma$~=~2.55~$\pm$~0.01, in close
agreement with the value given by Boller \etal (1996). Using either
two separate power laws or a broken power law model over the 
0.2--10~keV bandpass (fits 6 and 7 in Table~\ref{fits}) provides
statistically acceptable fits, although they are much worse than for the
blackbody parametrisation (or {\sc thCompfe}; see below). Thus the
soft X-ray spectrum shows intrinsic curvature.

The broad-band best-fit model (fit 4) gives a 0.2--10~keV flux of
7.85~$\times 10^{-12}$~erg~cm$^{-2}$~s$^{-1}$, which corresponds to a
luminosity of 3.35~$\times 10^{43}$~erg~s$^{-1}$; $\sim$~30~per~cent
of this luminosity is due to the multiple blackbody components used to
model the soft excess.  The 2--10~keV luminosity was measured to be
9.67~$\times 10^{42}$~erg~s$^{-1}$. Over the {\it ROSAT} band, the
power law model gave a luminosity very similar to the value given in Boller
\etal (1996) but about half that seen in an earlier {\it ROSAT}
observation (Boller \etal 1992).

\subsubsection{Variability Analysis}
\label{sec:specanal_epic_var}

\begin{figure}
\begin{center}
\includegraphics[clip,width=0.7\linewidth,angle=-90]{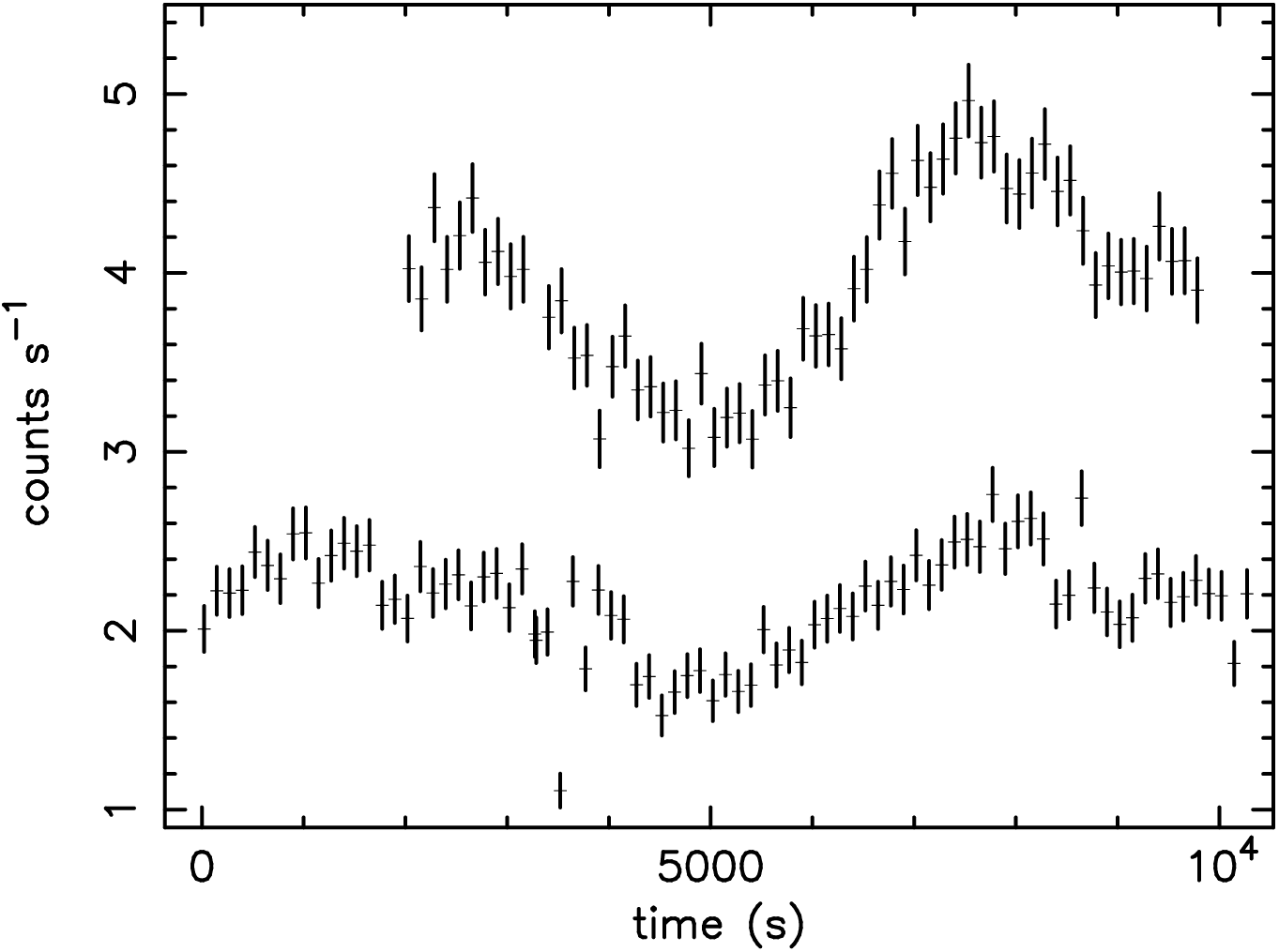}
\includegraphics[clip,width=0.35\linewidth,angle=-90]{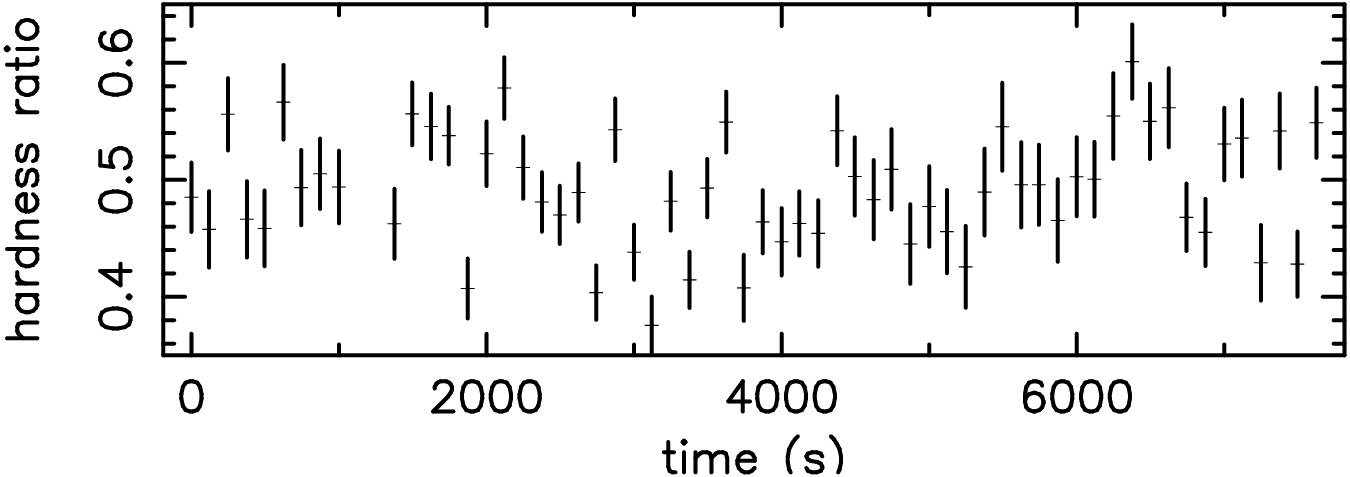}
\caption{Lightcurves showing the 0.2--10~keV continuum variability of
Mrk~896 over the 10~ks observation, using 125 second bins. The PN
camera was switched on approximately 2400 seconds after the MOS
instruments. The lower panel shows the hardness ratio between the
0.2--1 and 1--10~keV bands.}
\label{lc}
\end{center}
\end{figure}

Figure~\ref{lc} shows the MOS and PN 0.2--10~keV lightcurves for
Mrk~896 during the {\it XMM-Newton} obervation. There is a
$\pm$20-25~per~cent variation in count rate over a few thousand
seconds, which is consistent with other NLS1s (e.g., Boller, Brandt \&
Fink 1996). Boller \etal (1996) saw a variation by a factor of 1.9 in
just under 300~ks; here, we find a much faster variation, of
$\sim$~1.5 in 3~ks, although even this more rapid change only
corresponds to a radiative efficiency value of $\sim$~0.3~per~cent (Fabian 1979).
Over this short observation, we find no evidence for a difference in
the variability amplitude when comparing the 0.2--1~keV and 1--10~keV
bands; the calculated hardness ratio is fully consistent with being 
constant throughout the observation (lower panel of
Figure~\ref{lc}). Using a cross-correlation function, no significant
time delay between the hard and soft bands is found, with an upper
limit of $<$~270~seconds. Assuming a black hole mass of 10$^6$
$\msun$ (see section~\ref{sec:specanal_epic_comp}), this sets a limit of $<$~27~Schwarzschild radii for the size
of the emitting region.

\subsubsection{Comptonisation fits}
\label{sec:specanal_epic_comp}

Although modelling the spectrum with a power law and BB components
produces a good fit, it has no physical basis. Using the relationship
between the temperature of the accretion disc and the black hole mass
(e.g., Peterson 1997)

$$T(r) \sim 6.3 \times 10^{5} \left(\frac{\dot{M}}{\dot{M}_{Edd}}\right)^{1/4}M_{8}^{-1/4}\left(\frac{r}{R_{sch}}\right)^{-3/4} K$$

%\begin{displaymath}
%T(r) \sim
%6.3\times10^{5}\left(\frac{\dot{M}}{\dot{M}_{Edd}}\right)^{1/4}M_{8}^{-1/4}}\left(\frac{r}{R_{sch}}\right)^{-3/4} K
%\end{displaymath}

(where $\dot{M}$ is the mass accretion rate, $\dot{M}_{Edd}$
is the Eddington accretion rate, M$_{8}$ signifies the mass of the central black hole in units of
10$^{8}$ $\msun$ and R$_{sch}$ is the Schwarzschild radius), the
temperature expected for the inner-most part of the accretion disc
(r~=~3~R$_{sch}$), for 10$^6$ $\msun$ black hole radiating at
0.5$\dot{M}_{Edd}$ is $\sim$~60~eV. [The black hole mass was estimated by assuming a bolometric luminosity of ~$\sim$~10~$\times$ the 0.2--10~keV X-ray value and taking this to be the Eddington limit. This, together with the assumed accretion rate, is an acceptable value, since NLS1s are thought to be low mass systems with high
mass-accretion rates (Pounds, Done \& Osborne 1995).]
This is an upper limit to the accretion disc temperature, since areas
more distant from the black hole will be radiating at lower values. It can be seen that
the BB of $\sim$~80~eV is broadly consistent with this value, while the 200~eV component
could not be produced through thermal radiation from the disc.

A more realistic model
involves Comptonisation: soft photons from the accretion disc are
up-scattered by hot, thermal electrons, possibly located in a corona
above the disc. A two-temperature distribution leads to the formation
of both the soft excess and the harder power law slope. To determine
whether Comptonisation could be used to explain the Mrk~896 spectrum,
the {\sc Xspec} model {\sc comptt} was used. Initially, just the
soft excess was modelled in this fashion; i.e., the power law
component was used for the higher energy part of the spectrum, with
the iron line fixed as in Fit 2. This
led to a very good fit ($\chi^{2}_{\nu}$~=~704/715), for
$\Gamma$~=~2.06~$\pm$~0.04, together with an input photon temperature
of (67~$\pm$~5)~eV and a Comptonised component of
kT~=~(0.47~$\pm$~0.23)~keV and $\tau$~=~6.7~$\pm$~2.3.

When using a second {\sc comptt} component to replace the power law,
it was found that the temperature for the hotter
distribution could not be very well constrained. This is due to the
fact that the exponential cut-off produced in the Comptonisation fit
lies outside the {\it XMM} energy band-pass and cannot, therefore, be
easily determined. However, a statistically good fit is obtained when
the temperature is fixed at 200~keV; the
model parameters are given as Fit 8 in Table~\ref{compfits} and,
again, the iron line was kept as in Fit 2. It must be noted that, because of fixing the hotter kT, the error on the optical depth
appears small.

\begin{figure}
\begin{center}
\includegraphics[clip,width=0.7\linewidth,angle=-90]{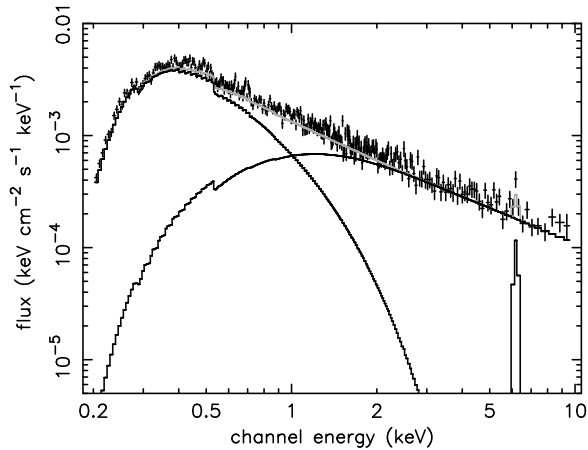}
\caption{An unfolded plot of the best fit {\sc thCompfe} model; again,
only the PN data are shown. The model consists of disc photons
at $\sim$~60~eV Comptonised by two electron distributions, at
$\sim$~0.3 and 200~keV respectively.}
\label{thcomp}
\end{center}
\end{figure}

There are many theories for the production of the X-ray spectra
through Comptonisation (and others which do not invoke Comptonising
distributions). When modelling the spectrum of Mrk~896 using {\sc
comptt}, it was assumed that cool disc photons were Comptonised by one
of two temperature distributions. Although this is a possible model,
it is more likely that there is a `layered' structure, such that most
of the thermal photons are first Comptonised by the cooler of the
electron distributions, forming the photons we see as the broad soft
excess; following this interaction, some of these photons will be
further Comptonised by the hotter electrons (likely to be produced
through magnetic reconnection above the accretion disc; this electron
distribution may be non-thermal), producing the power law tail seen at
higher energies. If this is indeed the case, then the input photon
temperature for the hotter Comptonisation should be that produced by
the first Comptonisation stage. This is the situation assumed when
fitting with {\sc thCompfe}, an alternative Comptonisation model
({\.Z}ycki, Done \& Smith 1999) which takes into account the roll-off
of the power law at lower energies due to the input photon
distribution. In this respect, it is a more self-consistent model, and
tends to give better-constrained results. Using {\sc thCompfe}
produces an excellent fit, with $\chi^{2}_{\nu}$ of 704/715 (Fit 9 in
Table~\ref{compfits}). The temperature of the hotter distribution was
fixed to 200~keV, but the cooler electron temperature was allowed to
float. Figure~\ref{thcomp} shows the spectrum. It
should be noted that assuming the same input photon population for
both Comptonising distributions also produces as good a fit, although
the temperature of the electrons producing the soft excess photons is
not well constrained. Overall the Comptonisation models provide as good
a fit to the data as the BB model since they reproduce the spectral
curvature better than multiple power law models. Using the F-test, a value of F~$>$~18 is obtained from the Comptonisation model over the power law fits of the soft excess; this corresponds to an improvement of $>$~99~per~cent.

\begin{table*}
\begin{center}
\caption{} 
\label{compfits}
Comptonisation fits to the broad-band {\it XMM} data.\\
$^{t}$ tied to cooler component
$^{f}$ frozen
\begin{tabular}{p{0.3truecm}p{1.3truecm}p{1.8truecm}p{1.8truecm}p{1.8truecm}p{1.8truecm}p{1.8truecm}p{1.8truecm}p{1.5truecm}}
\hline
 &  &   \multicolumn{3}{c} {\sc cooler comptonising component}  &  \multicolumn{3}{c}{\sc
 hotter comptonising  component}\\
& \\
Fit & Model & input photons & Compt. temp. & optical depth & input photons & Compt. temp. & optical
depth & $\chi^{2}$/dof\\
 &  & (eV) & (keV)& & (eV) & (keV) \\
\hline
8 & {\sc comptt} & 66~$\pm$~5 & 0.54~$\pm$~0.34 & 6.2~$\pm$~3.1 &
66$^{t}$  & 200$^{f}$  & 0.14~$\pm$~0.01 & 704/715\\
9 & {\sc thCompfe} & 58~$\pm$~5 & 0.28~$\pm$~0.18 & 13.8~$\pm$~6.3 &
280$^{t}$ & 200$^{f}$ & 0.77~$\pm$~0.03 & 704/715\\
\hline
\end{tabular}
\end{center}
\end{table*}

\subsubsection{Ionised disc fit}
\label{sec:specanal_epic_xion}
An ionised accretion disc reflection model, described by Ballantyne \etal 2001,
was also used, initially to fit the 2--10~keV energy range. The
best fit obtained was found to consist of a power law component, of
$\Gamma$~=~2.04~$\pm$~0.04, with an ionisation parameter of
1.37~$\pm$~0.35 erg~cm~s$^{-1}$ (i.e., only weakly ionised) and a reflection component,
R~=~0.52~$^{+0.55}_{-0.10}$, where R~=~$\Omega$/2$\pi$. This gave a
reduced $\chi^{2}$ value of 250/266, which is not significantly different
than the value of 246/263 for a Gaussian fit. 

Ionised reflection may contribute to the soft X-ray curvature observed in
broad-band AGN X-ray spectra, often removing the need for multiple
blackbodies (e.g., for Mrk~205, Reeves \etal 2001; Mrk~359, O'Brien \etal
2001; Mrk~509, Pounds \etal 2001). Even upon including the ionised
reflection model in the fits, and allowing the ionisation parameter to
vary, two BB components (with temperatures very
similar to the fit without the ionised reflector) were still required
to fit the broad-band Mrk~896
spectrum ($\chi^{2}_{\nu}$~=~716/717); however, the normalisations of the
blackbody components were reduced by approximately 25~per~cent in strength.
It is clear, therefore, that not {\it all} of the soft excess in this object
can be explained purely through the reprocessing of the primary, hard X-ray
component by an ionised disc.

\subsection{RGS data}
\label{specanal_rgs}

The RGS instrument allows a more detailed investigation of the soft
X-ray spectrum. In an earlier paper (O'Brien \etal 2001), the NLS1
Mrk~359 was analysed. The RGS data for this object were found to show
an absorption trough, possibly corresponding to Fe~M ions (e.g., Sako \etal 2001), together
with emission lines from O~{\sc viii} Ly$\alpha$ and the Ne~{\sc ix}
and O~{\sc vii} triplets. The RGS spectrum of Mrk~896 shows a weak absorption trough around
16.9~\AA~(rest frame), which could correspond to an Fe~M feature, and
a small peak around the energy expected for O~{\sc viii} Ly$\alpha$
(rest frame wavelength of 18.9~\AA). However, fitting a similar model
to that used for Mrk~359 provides only equivalent-width upper limits
for these spectral features of EW~$<$~18.4~eV and EW~$<$~12~eV
respectively. Similarly, the respective combined upper limits for the
Ne~{\sc ix} ($\lambda_{rest}$~=~13.6~\AA) and O~{\sc vii}
($\lambda_{rest}$~=~22.0~\AA) triplets are EW~$<$~11.6~eV and
$<$10.3~eV. There are also no strong absorption edges within the data, with $\tau$(O~{\sc vii})~$<$~0.35 and $\tau$(O~{\sc viii})~$<$~0.16 (at the 90~per~cent confidence level). 

The lack of strong spectral features in the soft energy
region implies that the observed soft excess is not dominated by  a
blend of soft X-ray lines (e.g., Turner \etal 1991). The shape of the spectrum is also very different from MCG$-$6$-$30$-$15 and Mrk~766, where relativisitc lines were used to fit the soft excess by Branduardi-Raymont \etal (2001).

\subsection{OM data}
\label{specanal_om}

\begin{figure}
\begin{center}
\includegraphics[clip,width=0.7\linewidth,angle=-90]{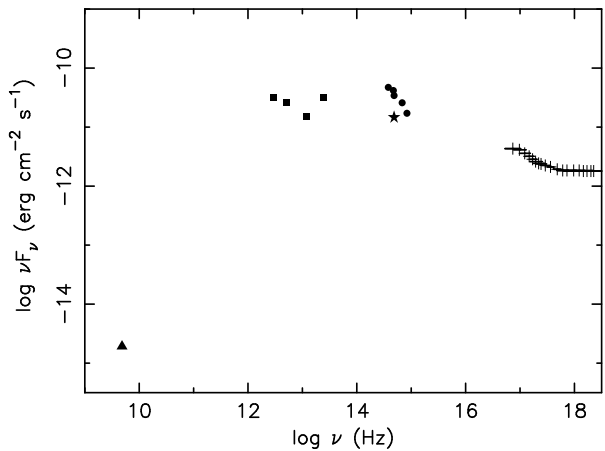}
\caption{A Spectral Energy Distribution of Mrk 896, showing the radio (triangle),
IR (squares), 
optical (circles and star) and X-ray (crosses) measurements.}
\label{sed}
\end{center}
\end{figure}

Mrk~896 was observed using the V-band filter of the Optical Monitor
only. A magnitude of m$_{V}$~=~15.28~$\pm$~0.01 was obtained,
corresponding to a flux of 2.735~mJy. The OM uses a small aperture
size of six arcsec. Winkler (1997), however, used a twenty arcsec
aperture, which includes more of the host galaxy, so obtaining a brighter magnitude of $\sim$~14.4.
Figure~\ref{sed} shows the position of the optical points (from
Winkler 1997 and the OM -- circle and star markers respectively) in relation to the X-ray data, together with
the radio measurement obtained from Bicay \etal (1995) and four IR
data points, obtained from the IRAS Faint Source Catalogue, version
2.0 (Moshir
\etal 1990). 

\section{Discussion}
\label{sec:disc}

A simple analysis of the broad-band {\it XMM-Newton} spectrum of Mrk~896 reveals
a soft excess lying above a power law, fitted over the 2--10~keV band,
as is typical for observations of AGN (e.g., Pounds \& Reeves 2002). The lack of strong spectral
features in the RGS spectrum appears to rule out the possibility that
the soft excess could be due to the blend of soft X-ray lines. It is found that a two-temperature
Comptonisation model, where photons from the accretion disc undergo
inverse-Compton scattering with hot electrons to produce the soft
excess and hard `power law' respectively, fits the spectrum very well.
The input
photons are $\sim$~60~eV (appropriate for emission from the inner-most
regions of the accretion disc) and the Comptonising distributions have
temperatures of kT~$\sim$~0.3 and 200~keV.
It is possible that the input photons for the hotter electron
distribution are those which have been previously Comptonised by the
soft-excess-producing population; however, the geometry of the inner
regions of active galactic nuclei is unknown. It could be that
photons emitted from the accretion disc are {\it directly}
Comptonised to form the harder power law slope. Alternatively, the hotter electrons may be
a non-thermal distribution (see Coppi 1999 and Vaughan \etal 2002 for details of hybrid
thermal/non-thermal plasmas).

Considering the temperatures found for the blackbody fit, the cooler
of the two components ($\sim$~85~eV) is consistent with being the 
high-energy tail of the big blue bump, while the hotter, $\sim$~225~eV
BB corresponds to the temperature of the soft-excess-producing
Comptonising component. (The power law replaces the higher-temperature Comptonised term.)

It should be noted that it is not possible statistically to differentiate between the blackbody and Comptonisation models fitted to these data.

A neutral Fe~K$\alpha$ emission line is also found. Although an ionised accretion disc model
can be invoked to explain the slight broadening of the iron emission
line, this does not account for the entire curvature at lower
energies; that is, the soft excess cannot arise simply through the 
reprocessing of the hard X-ray continuum. The line is, however, consistent with being a narrow feature; it is
becoming increasingly apparent that almost all broad-line AGN, below a
luminosity of $\sim$~10$^{45}$~erg~s$^{-1}$, show this feature, with
the lower luminosity objects having the stronger lines.  It is thought
that this narrow line is formed through fluorescence in 
distant, cool matter, possibly the BLR or putative molecular torus
(e.g. Ghisellini, Haardt \& Matt 1994), with higher luminosity objects
having smaller covering factors (O'Brien \etal 2001; Pounds \& Reeves 2002).

\section{ACKNOWLEDGMENTS}
The work in this paper is based on observations with {\it
XMM-Newton}, an ESA
science mission, with instruments and contributions directly funded by
ESA and NASA. The authors would like to thank the EPIC Consortium for all their work during the calibration phase, 
and the SOC and SSC teams for making the observation and analysis
possible. 
This research has made use of the NASA/IPAC Extragalactic
Database (NED), which is operated by the Jet Propulsion Laboratory,
California Institute of Technology, under contract with the National
Aeronautics and Space Administation.
Support from a PPARC studentship and the Leverhulme Trust is
acknowledged by KLP and JNR respectively.

\end{document}